\documentclass[12pt,fleqn]{article}
\usepackage{amsfonts,ulem,url}
\usepackage{amssymb}
\usepackage{amsmath}
\usepackage{amstext,verbatim,graphicx,ifthen,multirow,amsthm}
\usepackage{bm}
\usepackage{natbib}
\usepackage[bf]{caption}
\usepackage{setspace}

\renewcommand{\mathbf}{\boldsymbol}
\renewcommand{\thepage}{}
\renewcommand{\appendix}{\footnotesize\parindent 0cm\setcounter{equation}{0}
\renewcommand{\theequation}{A.\arabic{equation}}
\setcounter{lemma}{0}\renewcommand{\thelemma}{A.\arabic{lemma}}}

\newcommand{\adopt}{\mathbf{1}_{A_i\leq T}}

\newcommand{\bba}{\mathbf{a}}
\newcommand{\bfe}{\mathbf{1}}

\newcommand{\ba}{\mathbf{A}}
\newcommand{\bx}{\mathbf{X}}
\newcommand{\by}{\mathbf{Y}}
\newcommand{\bw}{\mathbf{W}}
\newcommand{\byp}{\mathbf{Y}^p}
\newcommand{\bypi}{\mathbf{Y}^p_i}

\newcommand{\cwit}{\dot{W}_{it}}

\newcommand{\did}{{\rm did}}

\newcommand{\htdid}{\hat{\tau}_{\mathrm{did}}}
\newcommand{\hvdid}{\widehat{\mathbb{V}}_{\mathrm{did}}}

\newcommand{\infone}{\infty 1}
\newcommand{\indep}{\perp\!\!\!\perp}

\newcommand{\minn}{{-}}
\newcommand{\mbf}{\mathbf{1}}
\newcommand{\mme}{\mathbb{E}}
\newcommand{\mmt}{\mathbb{T}}
\newcommand{\mma}{\mathbb{A}}
\newcommand{\mmv}{\mathbb{V}}
\newcommand{\mmc}{\mathbb{C}}

\newcommand{\ow}{\overline{W}}

\newcommand{\oy}{\overline{Y}}

\newcommand{\plus}{{+}}
\newcommand{\pr}{{\rm pr}}

\newcommand{\sumaap}{\sum_{a,a':a\neq a'}}

\newcommand{\tauinfone}{\tau_{t,\infone}}

\newtheorem{comments}{Comment}

\newtheorem{assumption}{Assumption}
\newtheorem{theorem}{Theorem}
\newtheorem{lemma}{Lemma}

\def\monthname{\ifcase\month\or
  January\or February\or March\or April\or May\or June\or July\or
  August\or September\or October\or November\or December\fi}
\numberwithin{equation}{section}

\def\monthname{\ifcase\month\or
January\or February\or March\or April\or May\or June\or
July\or August\or September\or October\or November\or December\fi}
\renewcommand{\appendix}{\small\parindent 0cm\setcounter{equation}{0}
\renewcommand{\theequation}{A.\arabic{equation}}
\setcounter{lemma}{0}\renewcommand{\thelemma}{A.\arabic{lemma}}
\setcounter{theorem}{0}\renewcommand{\thetheorem}{A.\arabic{theorem}}}
\begin{document}

\title{\textbf{Design-based Analysis in Difference-In-Differences Settings with  Staggered Adoption}\thanks{{\small We are
grateful for comments by participations in the conference in honor of Gary Chamberlain at Harvard in May 2018, and in particular by Gary Chamberlain. Gary's insights over the years have greatly affected our thinking on these problems. 
We also wish to thank  Sylvia Kloskin and Michael Pollmann for superb research assistance. This research was generously supported by ONR grant N00014-17-1-2131. }} }
\author{Susan Athey\thanks{{\small Professor of
Economics, Graduate School of Business, Stanford University, and NBER,
athey@stanford.edu. }}  \and Guido W. Imbens\thanks{{\small Professor of
Economics,
Graduate School of Business, Stanford University, SIEPR, and NBER,
imbens@stanford.edu.}} }
\date{ Current version \ifcase\month\or
January\or February\or March\or April\or May\or June\or
July\or August\or September\or October\or November\or December\fi \ \number%
\year\ \  }
\maketitle\thispagestyle{empty}

\begin{abstract}
\singlespacing
\noindent 
In this paper we study estimation of and inference for average treatment effects in a setting with panel data. We focus on the setting where  units, e.g., individuals, firms, or states, adopt the policy or treatment of interest at a particular point in time, and then remain exposed to this treatment at all times afterwards. We take a design perspective where we investigate the properties of estimators and procedures given assumptions on the assignment process. We show that under random assignment of the adoption date the standard Difference-In-Differences estimator is is an unbiased estimator of a particular weighted average causal effect. We characterize the properties of this estimand, and show that the standard variance estimator is conservative.
\end{abstract}


\begin{center}
\end{center}



\baselineskip=20pt\newpage
\setcounter{page}{1}
\renewcommand{\thepage}{\arabic{page}}
\renewcommand{\theequation}{\arabic{section}.\arabic{equation}}


\section{Introduction}

In this paper we study estimation of and inference for average treatment effects in a setting with panel data. We focus on the setting where  units, e.g., individuals, firms, or states, adopt the policy or treatment of interest at a particular point in time, and then remain exposed to this treatment at all times afterwards. The {adoption date} at which units are first exposed to the policy may, but need not, vary by unit. We refer to this as a {\it staggered adoption design} (SAD), such designs  are sometimes also referred to as event study designs. An early example is \citet{athey1998empirical} where adoption of an enhanced 911 technology by counties occurs over time, with the adoption date varying by county. This setting is a special case of the general Difference-In-Differences (DID)  set up (e.g., \citet{cardmariel, meyer1995workers, angristpischke,  angristkruegerstrategies,  Abadie2010, borusyak2016revisiting, atheyimbenscic, cardkrueger1, freyaldenhoven2018pre, de2018two, abadie2005semiparametric}) where, at least in principle, units can switch back and forth between being exposed or not to  the treatment.
In this SAD setting we are concerned with identification issues as well as estimation and inference. In contrast to most of the DID literature, e.g.,  \citet{Bertrand2004did, shah1977inference,  conley2011inference,    donald2007inference, stock2008heteroskedasticity, manuel1987computing, arellano2003panel, abraham2018estimating, wooldridge2010econometric, de2017fuzzy, de2018two}, we  take a {\it design-based}  perspective where the stochastic nature and   properties of the estimators arises from the stochastic nature of the assignment of the treatments, rather than  a {\it sampling-based} perspective where the uncertainty arises from the random sampling of  units from a large population. Such a design perspective is common in the analysis of randomized experiments, e.g., \citet{neyman1923, rosenbaum_book, rosenbaum2017observation}. See also \citet{aronow2016, abadieathey2, abadie2017sampling} for this approach in  cross-section regression settings. This perspective is particularly attractive in the current setting when the sample comprises the entire population, e.g., all states of the US, or all countries of the world. Our critical assumptions involve restrictions on the assignment process as well as exclusion restrictions, but in general do not involve functional form assumptions. Commonly made common trend assumptions (\citet{de2018two, abraham2018estimating}) follow from some of our assumptions, but are not the starting point.

As in \citet{abraham2018estimating}
we  set up the problem with the adoption date, rather than the actual exposure to the intervention, as the basic treatment defining the potential outcomes. We consider assumptions under which this discrete multivalued  treatment (the adoption date) can be reduced to a binary one, defined as the indicator whether or not the treatment has already been adopted. We then investigate the interpretation of the standard DID estimator under assumptions about the assignment of the adoption date and under various exclusion restrictions. We show that under a random adoption date assumption, the standard DID estimator
can be interpreted as the weighted average of several types of causal effects; within our framework, these concern the impact of different types of changes in the adoption date of the units.  We also consider design-based inference for this estimand. 
We derive the exact variance of the DID estimator in this setting.
We show that under a random adoption date assumption the standard  Liang-Zeger (LZ) variance estimator (\citet{liang1986longitudinal, Bertrand2004did}), or the clustered bootstrap, are  conservative. For this case we propose an improved (but still conservative) variance estimator. 

Our paper is most closely relateds  to a very interesting set of recent papers on DID methods that explicitly  focus on issues with heterogenous treatment effects (\citet{abraham2018estimating, de2018two, han2018identification, goodman2017difference, callaway2018difference, hull2018estimating, strezhnev, imai2018matching, hazlett2018trajectory}, and \citet{borusyak2016revisiting}). 
Among other things these papers derive interpretations of the DID estimator as weighted averages of causal effects and bias terms under various assumptions.  In many cases they find that these interpretations involve weighted averages of basic average causal effects with potentially negative weights and propose alternative estimators that do not involve negative weights.

\section{Set Up}

Using the potential outcome framework for causal inference, we consider a setting with
a population of $N$ units.  Each of these $N$ units are characterized by a set of potential outcomes in $T$ periods for $T+1$ treatment levels, $Y_{it}(a)$. Here  $i\in\{1,\ldots,N\}$ indexes the units, $t\in\mmt=\{1,\ldots,T\}$ indexes the time periods, and the argument of the potential outcome function $Y_{it}(\cdot)$, $a\in\mma=\mmt\cup\{\infty\}=\{1,\ldots,T,\infty\}$ indexes the discrete treatment, the date that the binary policy was first adopted by a unit. Units can adopt the policy at any of the time periods $1,\ldots,T$, or not adopt the policy at all during the period of observation, in which case we code the adoption date  as $\infty$. Once a unit adopts the treatment, it remains exposed to the treatment for all periods afterwards.
This set up is like that in  \citet{abraham2018estimating, hazlett2018trajectory}, and in contrast to most of the DID literature where the binary indicator whether a unit is exposed to the treatment in the current period indexes the potential outcomes.
 We observe for each unit in the population
the adoption date $A_{i}\in\mma$ and the sequence of $T$ realized outcomes, $Y_{it}$, for $t\in\mmt$, where 
\[Y_{it}\equiv Y_{it}(A_{i}),\]
 is the realized outcome for unit $i$ at time $t$. We may also observe pre-treatment characteristics, denoted by the $K$-component vector $X_i$, although for most of the discussion we abstract from their presence. Let $\by$, $\ba$, and $\bx$ denote the $N\times T$, $N\times 1$, and $N\times K$ matrices with typical elements $Y_{it}$, $A_i$, and $X_{ik}$ respectively.
Implicitly we have already made a sutva-type assumption (\citet{rubin1978bayesian, imbens2015causal})  that units are not affected by the treatments (adoption dates) for other units.
Our design-based analysis views the potential outcomes $Y_{it}(a)$ as deterministic, and only the adoption dates $A_i$, as well as functions thereof such as the realized outcomes as stochastic. Distributions of estimators will be fully determined by the adoption date distribution, with the number of units $N$ and the number of time periods $T$ fixed, unless explicitly stated otherwise. Following the literature we refer to this as a randomization, or designed-based, distribution (\citet{rosenbaum2017observation, imbens2015causal, abadie2017sampling}), as opposed to a sampling-based distribution.

In many cases the units themselves are clusters of units of  a lower level of aggregation. For example, the units may be states, and the outcomes could be averages of outcomes for individuals in that state, possibly of samples drawn from subpopulations from these states. 
In such cases $N$ and $T$ may be as small as 2, although in many of the cases we consider $N$ will be at least moderately large.
This distinction between cases where $Y_{it}$ is itself an average over basic units or not, affects some, but not all, of the formal statistical analyses. It may make some of the assumptions more plausible, and it may affect the inference, especially if individual level outcomes and covariates are available.

Define $W(a,t)=\bfe_{a\leq t}$ to be the binary indicator for the adoption date $a$ preceeding $t$, and define $W_{it}$ to be the indicator for the the policy having been adopted by unit $i$ prior to, or at, time $t$:
\[ W_{it}\equiv W(A_i,t)= \bfe_{A_i\leq t},\]
so that the $N\times T$ matrix $\bw$ with typical element $W_{it}$ has the form:
\[ \bw_{N\times T}=\left(
\begin{array}{ccccccr}
 0  &  0  &  0  &  0   & \dots &  0  & {\rm (never\ adopter)}\\
 0  &  0   &  0  &  0    & \dots &  1   & {\rm (late\ adopter)}\\
 0  &  0  &  0  &  0    & \dots &  1   \\
 0  &  0  & 1  &  1    & \dots &  1   \\
 0  &  0  &  1  &  1    & \dots &  1  &\ \ \  {\rm (medium\ adopter)} \\
\vdots   &  \vdots   & \vdots & \vdots &\ddots &\vdots \\
 0  &  1  &  1  &  1    & \dots &  1  & {\rm (early\ adopter)}  \\
\end{array}
\right)
\]
Let $N_a\equiv \sum_{i=1}^N \bfe_{A_i=a}$ be the number of units in the sample with adoption date $a$, and
define $\pi_a\equiv N_a/N$, for $a\in\mma$, as the fraction of units with adoption date equal to $a$,
and $\Pi_t\equiv \sum_{s=1}^t \pi_s$, for $t\in\mmt$, as the fraction of units with an adoption date on or prior to $t$.

Also define $\oy_t(a)$ to be the population average of the potential outcome in period $t$ for adoption date $a$:
\[ \oy_t(a)\equiv \frac{1}{N}\sum_{i=1}^N Y_{it}(a),\hskip1cm {\rm for}\ t\in\mmt, a\in\mma.\]
Define the average causal effect of adoption date $a'$ relative to $a$, on the outcome in period $t$, as
\[ \tau_{t,aa'}\equiv \oy_t(a')-\oy_t(a)=\frac{1}{N}\sum_{i=1}^N \Bigl\{Y_{it}(a')-Y_{it}(a)\Bigr\}.
\]
  \citet{abraham2018estimating} focus on slighlty different building blocks, what they call $CATT_{a,t}$, which,  for $0\leq t\leq T-a$, are the super-population equivalent of
  $(1/N_a)\sum_{i|A_i=a} \{Y_{ia+t}(a)-Y_{ia+t}(\infty)\}$.
The average  causal effects $\tau_{t,aa'}$ are the building blocks of many of the estimands we consider later.
A particularly interesting average effect  is
\[ \tauinfone=\frac{1}{N}\sum_{i=1}^N \Bigl(Y_{it}(1)-Y_{it}(\infty)\Bigr),
\]
the average effect of switching the entire population from never adopting the policy ($a=\infty$), to adopting the policy in the first period ($a=1$). Formally there is nothing special about the particular average effect $\tauinfone$ relative to any other $\tau_{t,aa'}$, but $\tauinfone$ will be useful as a benchmark. Part of the reason is that for all $t$ and $i$ the comparison $Y_{it}(1)-Y_{it}(\infty)$  is between potential outcomes for adoption prior to or at time $t$ (namely adoption date $a=1$) and potential outcomes for  adoption later than $t$ (namely, never adopting, $a=\infty$). In contrast, any other average effect $\tau_{t,aa'}$ will for some $t$ involve comparing potential outcomes neither of which correspond to having adopted the treatment yet, or comparing potential outcomes both of which correspond to having adopted the treatment already. Therefore, $\tauinfone$ reflects more on the effect of having adopted the policy than any other $\tau_{t,aa'}$.

\section{Assumptions}

We consider three sets of assumptions. The first set, containing only a single assumption, is about the design, that is, the assignment of the treatment, here the adoption date, conditional on the potential outcomes and possibly pretreatment variables.
We refer to this as a design assumption because it can be guaranteed by design.
The second set of assumptions is about the potential outcomes, and rules out the presence of certain treatment effects.
These exclusion restrictions are substantive assumptions, and they cannot be guaranteed by design.
The third set of assumptions consists of four auxiliary assumptions, two  about homogeneneity of certain causal effects, one about sampling from a large population, and one about an outcome model in a large population.
The nature of these three sets of assumptions, and their plausibility, is very different, and it is in our view  useful to carefully distinguish between them.
The current literature often combines various parts of these assumptions implicitly in the notation used and in  assumptions about the statistical models for the realized outcomes.

\subsection{The Design Assumption}

The first assumption is about the assignment process for the adoption date $A_i$. 
Our starting point is to assume that the adoption date is completely random:
\begin{assumption}\label{assumption_completely_random}{\sc (Random Adoption Date)}
For some set of positive integers $N_a$, for $a\in\mma$,
\[ \pr(\ba=\bba)
=\left(
\frac{N!}{\prod_{a\in\mma} N_a!}
\right)^{-1},\]
for all $N$-vectors $\bba$ such that for all $a\in\mma$,
 $\sum_{i=1}^N \mathbf{1}_{a_i=a} =N_a$.
\end{assumption}
This assumption is obviously very strong. However, without additional assumptions that restrict either  the potential outcomes, or expand what we observe, for example by including pre-treatment variables or covariates, this assumption has no testable implications in a setting with exchangeable units.
\begin{lemma}\label{lemma:norestr} {\sc (No Testable Restrictions)} Suppose all units are exchangeable. Then
Assumption \ref{assumption_completely_random} has no testable implications for the joint distribution of $(\by,\ba)$.
\end{lemma}
\noindent All proofs are given in the Appendix.

Hence, if we wish to relax the assumptions, we need to bring in additional information. Such additional information can come in the form of pretreatment variables, that is, variables that are known not to be affected by the treatment. 
In that case we can relax the assumption by requiring only that the adoption date is completely random within subpopulations with the same values for the pre-treatment variables.
Additional information can also come in the form of limits on the treatment effects. The implications of such restrictions on the ability to relax the random adoption assumption is more complex, as discussed in more detail in Section \ref{excl}.

Under Assumption \ref{assumption_completely_random}
the marginal distribution of the adoption dates is fixed, and so also the fraction $\pi_a$ is fixed in the repated sampling thought experiment. This part of the set up is similar in spirit  to  fixing the number of treated units in the sample in a completely randomized experiment.
It is convenient for obtaining finite sample results. Note that it implies that the adoption dates for units $i$ and $j$ are not independent.
Note also that in the standard framework where the uncertainty arises solely from random sampling, this fraction does not remain constant in the repeated sampling thought experiment.

An important role is played by what we label  the {\it adjusted treatment}, adjusted for unit and time period averages:
\[ \cwit\equiv
W_{it}-\ow_{\cdot t}-\ow_{i\cdot}+\ow,\]
where $\ow_{\cdot t}$, $\ow_{i\cdot}$, and $\ow$ are averages over units, time periods, and both, respectively:
\[\ow_{\cdot t}\equiv \frac{1}{N}\sum_{i=1}^N W_{it}= \frac{1}{N}\sum_{i=1}^N\mbf_{A_i\leq t}=
\frac{1}{N}\sum_{i=1}^N \sum_{s\leq t}\mbf_{A_i=s}
=
\sum_{s\leq t}\frac{1}{N}\sum_{i=1}^N \mbf_{A_i=s}
=\sum_{s\leq t}\pi_s,
\]
\[ \ow_{i\cdot}\equiv \frac{1}{T}\sum_{t=1}^T W_{it}
=\adopt\frac{T+1-A_i}{T}, \]
and
\[ \ow\equiv \frac{1}{T}\sum_{t=1}^T \ow_{\cdot t}=\frac{1}{T}\sum_{t=1}^T \sum_{s\leq t} \pi_s=
\frac{1}{T}\sum_{t=1}^T (T+1-t)\pi_t,\]
where, with some minor abuse of notation, we adopt the convention that $a\mathbf{1}_{a\leq T}$ is zero if $a=\infty$.
Note that under Assumption \ref{assumption_completely_random}, $ow_{\cdot t}$ and $\ow$ are non-stochastic.
Using these representations  we can write the adjusted treatment indicator as
\[ \cwit=
g(t,A_i),\]
where
\begin{equation}\label{eq_g}
g(t,a)\equiv 
\left(\mathbf{1}_{a\leq t}-\sum_{s\leq t} \pi_s \right)
+\frac{1}{T}\left(a\mathbf{1}_{a\leq T}-\sum_{s=1}^T s\pi_s\right)
+\frac{T+1}{T}\left(\mathbf{1}_{a=\infty}-\pi_\infty\right)
.\end{equation}
Because the marginal distribution of $A_i$ is fixed under  Assumption  \ref{assumption_completely_random}, the sum $\sum_{i,t} \cwit^2$ is non-stochastic under this assumption, even though $\cwit$ and thus $\cwit^2$ are stochastic. This fact enables us to derive exact finite sample results for the standard DID estimator as discussed in Section \ref{section:four}. This is similar in spirit to the derivation of the exact variance for the estimator for the average treatment effect in completely randomized experiments when we fix the number of treated and controls.

\subsection{Exclusion Restrictions}\label{excl}

The next two assumptions concern the potential outcomes. Their formulation does not involve the assignment mechanism, that is, the distribution of the adoption date.
In essence these are  exclusion restrictions, assuming that particular causal effects are absent. Collectively these two assumptions imply that we can think of the treatment as a binary one, the only relevant component of the adoption date being whether a unit is exposed to the treatment at the time we measure the outcome. Versions of such assumptions are also considered in \citet{borusyak2016revisiting,  de2018two, abraham2018estimating, hazlett2018trajectory} and
 \citet{imai2016should}, where in the latter a graphical approach is taken in the spirit of the work by 
\citet{pearl}.

The first of the two assumptions, and likely the more plausible of the two in practice, rules out effects of future adoption dates on current outcomes. More precisely, it assumes that if the policy has not been adopted yet, the exact future date of the adoption has no causal effect  on potential outcomes for the current period. 
\begin{assumption}\label{assumption_no_anticipation}{\sc (No Anticipation)}
{\sc }
For all units $i$, all time periods $t$, and for all adoption dates $a$, such that $a>t$,
\[ Y_{it}(a)=Y_{it}(\infty).\]
\end{assumption}
We can also write this assumption as requiring that for all $(i,t,a)$,
\[ Y_{it}(a)=\bfe _{a\leq t} Y_{it}(a)+\bfe _{a> t} Y_{it}(\infty),
\hskip1cm {\rm or}\ \ \bfe _{a> t} \Bigl(Y_{it}(a)-Y_{it}(\infty)\Bigr)=0
,\] with the last representation showing most clearly how the assumption rules out certain causal effects.
Note that this assumption does not involve the adoption date, and so does not restrict the distribution of the adoption dates. Violations of this assumption may arise if the policy is anticipated prior to its implementation.

The next assumption is arguably much stronger. It asserts that for potential outcomes in period $t$ it does not matter how long the unit has been exposed to the treatment, only whether the unit is exposed at time $t$.
\begin{assumption}\label{assumption_exposure}{\sc (Invariance to History)}
{\sc }
For all units $i$, all time periods $t$, and for all adoption dates $a$, such that 
$a\leq t$,
\[ Y_{it}(a)=Y_{it}(1).\]
\end{assumption}
This assumption can also be written as
\[ Y_{it}(a)=\bfe _{a\leq t} Y_{it}(1)+\bfe _{a> t} Y_{it}(a)
,
\hskip1cm {\rm or}\ \ \bfe _{a\leq  t} \Bigl(Y_{it}(a)-Y_{it}(1)\Bigr)=0,\]
with again the last version of the assumption illustrating the exclusion restriction in this assumption.
Again, the assumption does not rule out any correlation between the potential outcomes and the adoption date, only that there is no causal effect of an early adoption versus a later adoption on the outcome in period $t$, as long as adoption occurred before or on  period $t$.

In general, this assumption is very strong. However, there are important cases where it may be more plausible. Suppose the units are clusters of individuals, where in each period we observe different sets of individuals. To be specific, suppose the the units are states, the time periods are years, and outcome is the employment rate for twenty-five year olds, and the treatment is the presence or absence of some regulation, say a subsidy for college tuition. In that case it may well be reasonable to assume that the educational choices for students graduating high school in a particular state depends on what the prevailing subsidy is, but much less on the presence of subsidies in previous years.

If both the exclusion restrictions, that is, both Assumptions \ref{assumption_no_anticipation} and \ref{assumption_exposure}, hold, 
then the potential outcome $Y_{it}(a)$ can be indexed by the binary indicator $W(a,t)=\bfe _{a\leq t}$:
\begin{lemma}\label{lemma_binary_treatment}{\sc (Binary Treatment)}
Suppose Assumptions \ref{assumption_no_anticipation} and \ref{assumption_exposure} hold. Then
for all units $i$, all time periods $t$ and adoption dates $a>a'$, $(i)$ 
\[ Y_{it}(a')-Y_{it}(a)=\bfe_{a'\leq t<a}\Bigl(Y_{it}(1)-Y_{it}(\infty)\Bigr),\]
so that,
\[ Y_{it}(a)
=Y_{it}(\infty)+
\bfe _{a\leq t}\Bigl(Y_{it}(1)-Y_{it}(\infty)\Bigr)
=\left\{\begin{array}{ll}Y_{it}(\infty)\hskip1cm &{\rm if} \ a\leq t\\
Y_{it}(1)\hskip1cm &{\rm otherwise,} 
\end{array}\right.
\]
and, for all time periods $t$, and adoption dates $a>a'$, $(ii)$
\[ \tau_{t,aa'}=\tau_{t,\infty 1} \bfe _{a'\leq t<a}
=\left\{\begin{array}{ll}
\tau_{t,\infty 1}\hskip1cm & {\rm if}\ a'\leq t<a,\\
0 & {\rm otherwise.}\end{array}
\right.
\]
\end{lemma}
If these two assumptions hold,  we can  therefore
simplify the notation for the potential outcomes and focus on $Y_{it}(1)$ and $Y_{it}(\infty)$.

Note that these two assumptions are substantive, and cannot be guaranteed by design.
This in contrast to the  Assumption \ref{assumption_completely_random}, which can be guaranteed by randomization of the adoption date.
It is also important to note that in many empirical studies  Assumptions 
 \ref{assumption_no_anticipation} and \ref{assumption_exposure} are  made,  often implicitly by writing a model for realized outcome $Y_{it}$ that depends solely on the contemporaneous treatment exposure $W_{it}$, and not on the actual adoption date $A_i$ or treatment exposure $W_{it'}$ in other periods $t'$. In the current discussion we want to be explicit about the fact that this restriction is an assumption, and that it does not automatically hold. Note that the assumption does not restrict the time series dependence between the potential outcomes.

It is trivial to see that without additional information, the exclusion restrictions in Assumptions  \ref{assumption_no_anticipation} and \ref{assumption_exposure} have no testable implications because they impose restrictions on pairs of potential outcomes that cannot be observed together. However, in combination with random assignment,  \ref{assumption_no_anticipation} and \ref{assumption_exposure}, there are testable implications as long as $T\geq 2$ and there is some variation in the adoption date.
\begin{lemma}\label{lemma:testable}
{\sc (Testable Restrictions from the Exclusion Restrictions)}
$(i)$ 
Assumptions  \ref{assumption_no_anticipation} and \ref{assumption_exposure} jointly  have no testable implications for the joint distribution of $(\by,\bw)$.
\\
$(ii)$ Suppose $T\geq 2$, and $\pi_2,\pi_\infty>0$. Then the combination of Assumptions  \ref{assumption_completely_random}--\ref{assumption_exposure} impose testable restrictions on the joint distribution of $(\by,\bw)$.
\end{lemma} 

\subsection{Auxiliary Assumptions}

In this section we consider four auxiliary assumptions that are convenient for some analyses, and in particular can have implications for the variance of specific estimators, but that are not essential in many cases. These assumptions are often made in empirical analyses without researchers explicitly discussing them.

The first of these assumptions assumes that the effect of adoption date $a'$, relative to adoption date $a$, on the outcome in period $t$, is the same for all units. 
\begin{assumption}\label{assumption_constant_treatment_effect}{\sc (Constant Treatment Effect Over Units)}
For all units $i,j$ and for all time periods $t$ and all adoption dates $a$ and $a'$
\[ Y_{it}(a)-Y_{it}(a')=Y_{jt}(a)-Y_{jt}(a').\]
\end{assumption}
The second assumption restricts the heterogeneity of the treatment effects  over time.
\begin{assumption}\label{assumption_constant_treatment_effect_time}{\sc (Constant Treatment Effect over Time)}
For all units $i$ and all time periods $t$ and $t'$
\[ Y_{it}(1)-Y_{it}(\infty)=Y_{it'}(1)-Y_{it'}(\infty).\]
\end{assumption}
 We only restrict the time variation for comparisons of the adoption dates 1 and $\infty$ because we typically use this assumption in combination with Assumptions \ref{assumption_no_anticipation} and \ref{assumption_exposure}. In that case we obtain a constant binary treatment effect set up, as summarized in the following Lemma.
\begin{lemma}
\label{lemma_did}{\sc (Binary Treatment and Constant Treatment Effects)}
Suppose Assumptions \ref{assumption_no_anticipation}-\ref{assumption_constant_treatment_effect_time} hold. 
Then for all $t$ and $a'<a$
\[ Y_{it}(a')-Y_{it}(a)=\mbf_{a'\leq t<a}\tau_{1\infty} .\]
\end{lemma}

The final assumption allows us to view the potential outcomes as random by postulating a large population from which the sample is drawn.
\begin{assumption}\label{assumption_random_sampling}{\sc (Random Sampling)}
The sample can be viewed as a random sampling from an infinitely large population, with joint distribution for $(A_i,Y_{it}(a),a\in\mma,t\in\mmt)$ denoted by $f(a,y_{1}(1),\ldots,y_T(\infty))$.
\end{assumption}
Under this assumption we can put additional structure on average potential outcomes. 
\begin{assumption}\label{assumption_additivity}{\sc (Additivity)}
\[\mme\left[ Y_{it}(\infty)\right]=\alpha_i+\beta_t.\]
\end{assumption}

\section{Difference-In-Differences Estimators: Interpretation and Inference}
\label{section:four}

In this section we consider the standard DID set up
(e.g.,  \citet{meyer1995workers,  Bertrand2004did, angristpischke, donald2007inference, de2018two}).
In the simplest setting with $N$ units and $T$ time periods, without additional covariates, the realized outcome in period $t$ for unit $i$ is modeled as
\begin{equation}\label{spec_did} Y_{it}=\alpha_i+\beta_t+\tau W_{it}+\varepsilon_{it}.\end{equation}
In this model there are unit effects $\alpha_i$ and time effects $\beta_t$, but both are additive with interactions between them ruled out. The effect of the treatment is implicitly assumed to be additive and constant across units and time periods.

We interpret the DID estimand under the randomized adoption date assumption, leading to a different setting from that considered in \citet{de2018two, abraham2018estimating, goodman2017difference}. We also derive its variance and
show that in general it is lower than the standard random-sampling based variance. Finally we propose a variance estimator that is is smaller than the regular variance estimators such as the Liang-Zeger and clustered bootstrap variance estimators.

\subsection{Difference-In-Differences Estimators}

Consider the least squares estimator for  $\tau$ based on the specification in (\ref{spec_did}): 
\[ 
\left(\htdid,\{\hat\alpha_i\}_{i=2}^N,\{\hat\beta_t\}_{t=1}^T\right)
=\arg\min_{\tau,\{\alpha_i\}_{i=2}^N,\{\beta_t\}_{t=1}^T}
\sum_{i=1}^N \sum_{t=1}^T 
\left(
Y_{it}-\alpha_i-\beta_t-\tau W_{it}\right)^2.\]
It is convenient to write $\htdid$ in terms of the adjusted treatment indicator $\cwit$  as
\[\htdid=\frac{\sum_{i,t} \cwit  Y_{it} }{\sum_{i,t}\cwit^2}.\]
The primary question of interest in this section concerns the properties of the estimator $\htdid$. This includes the interpretation of its expectation under various sets of assumptions, and its variance. Mostly we focus on exact properties in finite samples.

In order to interpret the expected value of $\htdid$  we  consider some intermediate objects.
Define, for all adoption dates $a\in\mma$, and all time periods $t\in\mmt$ the average of the outcome in period $t$ for units with adoption date $a$:
\[ \oy_{t,a}=\left\{
\begin{array}{ll}
\frac{1}{N_a} \sum_{i:A_i=a} Y_{it}\hskip1cm & {\rm if}\  N_a>0,\\
0 & {\rm otherwise.}\end{array}\right.\]
Under Assumption  \ref{assumption_completely_random} the stochastic properties of these averages are well-defined because   the $N_{a}$ are fixed over the randomization distribution. The averages are stochastic because the realized outcomes depend on the adoption date.
Define also the following two difference between outcome averages:
\[ \hat\tau_{t,aa'}
=\oy_{t,a'}-\oy_{t,a}
.\]
In general these differences do not have a causal interpretation. Such an interpretation  requires some assumptions, for example, on random assignment of the adoption date.

\noindent {\sc Example:} To facilitate the interpretation of some of the results it is useful to consider a special case where the results from completely randomized experiments directly apply. Suppose $\mmt=\{1,2\}$, and $\mma=\{2,\infty\}$, with a fraction $\pi=\pi_2=1-\pi_\infty$ adopting the policy in the second period.  Suppose also that $Y_{i1}(a)=0$ for all $i$ and $a$. Then the DID estimator is
\[ \htdid=\hat\tau_{2,2\infty}=\oy_{2,2}-\oy_{2,\infty}=\frac{1}{N_2}\sum_{i:A_i=2} Y_{i2}-\frac{1}{N_\infty}\sum_{i:A_i=\infty} Y_{i2},\]
the simple difference in means for the second period outcomes for adopters and non-adopters.
Under Assumption \ref{assumption_completely_random}, the standard results for the variance of the difference in means for a randomized experiments apply (e.g., \citet{neyman1923, imbens2015causal}), and the exact variance of $\htdid$ is, 
\[ \mmv(\htdid)=\frac{1}{N_2(N-1)}
\sum_{i=1}^N \left\{Y_{i2}(2)-\oy_2(2)\right\}^2 + \frac{1}{N_\infty(N-1)}
\sum_{i=1}^N \left\{Y_{i2}(\infty)-\oy_2(\infty)\right\}^2\]
\[\hskip2cm - \frac{1}{N(N-1)}
\sum_{i=1}^N \left\{\left(Y_{i2}(2)-\oy_2(2)\right)
-\left(Y_{i2}(\infty)-\oy_2(\infty)\right)
\right\}^2.\]
The standard Neyman estimator for this variance ignores the third term, and uses unbiased estimators for the first two terms, leading to:
\[ \hat\mmv(\htdid)=\frac{1}{N_2(N_2-1)}
\sum_{i:A_i=2} \left\{Y_{i2}-\oy_{2,2}\right\}^2
 + \frac{1}{N_\infty(N_\infty-1)}
\sum_{i:A_i=\infty} \left\{Y_{i2}-\oy_{2,\infty}\right\}^2.\]
$\square$

\subsection{The Interpretation of Difference-In-Differences Estimators}

The following weights play an important role in the interpretation of the DID estimand:
\begin{equation}
\label{eq:did1}
\gamma_{t,a}\equiv \frac{\pi_a g(t,a)}{\sum_{t'\in\mmt}\sum_{a'\in\mma}
\pi_{a'} g(t',a')^2}
,\hskip1cm \gamma_{t,\plus}\equiv \sum_{a\leq t} \gamma_{t,a},\hskip1cm {\rm and}\ \ \gamma_{t,\minn}\equiv \sum_{a> t} \gamma_{t,a}
,
\end{equation}
with $g(a,t)$ as defined in (\ref{eq_g}).
Note that these weights are non-stochastic, that is, fixed over the randomization distribution.

\noindent{\sc Example (ctd):}
Continuing the example with two periods and adoption in the second period or never, we have in that case
\[ \gamma_{t,a}=
\left\{
\begin{array}{ll}
0 \hskip1cm &{\rm if}\ (t,a)=(1,1),\\
0 \hskip1cm &{\rm if}\ (t,a)=(2,1),\\
-1 & {\rm if}\ (t,a)=(1,2),\\
1 & {\rm if}\ (t,a)=(2,2),\\
1 & {\rm if}\ (t,a)=(1,\infty),\\
-1 & {\rm if}\ (t,a)=(2,\infty),\\
\end{array}\right.
\hskip0.5cm
\gamma_{t,+}=
\left\{
\begin{array}{ll}
0 & {\rm if}\ t=1,\\
1 & {\rm if}\ t=2,
\end{array}\right.
\hskip0.5cm {\rm and}\ \ 
\gamma_{t,-}=
\left\{
\begin{array}{ll}
0 & {\rm if}\ t=1,\\
-1 & {\rm if}\ t=2.\\
\end{array}\right.
\]
$\square$

The weights $\gamma_{t,a}$ have some important properties,
\[
\sum_{t\in\mmt}\gamma_{t,\plus}=1
\hskip1cm 
\sum_{t\in\mmt}\gamma_{t,\minn}=-1,
\hskip1cm{\rm and}\ \ 
\sum_{t=1}^T\sum_{a\in\mma} \gamma_{t,a}=\sum_{t\in\mmt}\gamma_{t,\plus}
+\sum_{t\in\mmt}\gamma_{t,\minn}=
0. 
\]
Now we can state the first main result of the paper.
\begin{lemma}\label{lemma_did1}
We can write $\htdid$ as
\begin{equation}
\hat\tau_\did=\sum_{t\in\mmt}\sum_{a\in\mma} \gamma_{t,a}\oy_{t,a}
=\sum_{t\in\mmt} \gamma_{t,\plus} \hat\tau_{t,\infty 1}
+\sum_{t\in\mmt}\sum_{a> t} \gamma_{t,a} \hat\tau_{t,\infty a}
-\sum_{t\in\mmt}\sum_{a\leq t} \gamma_{t,a} \hat\tau_{t,a 1}
.
\end{equation}
\end{lemma}
\begin{comments}\rm Alternative characterizations of the DID estimator or estimand as a weighted average of potentially causal comparisons are presented in 
 \citet{abraham2018estimating, de2018two, han2018identification, goodman2017difference}, and \citet{borusyak2016revisiting}). The characterizations differ in terms of the building blocks that are used in the representation and the assumptions made.
Like our representation, the representation in   \citet{abraham2018estimating}   is in terms of average causal effects of different adoption dates, but it imposes no-anticipation.   \citet{goodman2017difference} presents the DID estimator in terms of basic two-group DID estimators.
Like our representation, the \citet{goodman2017difference} is mechanical and does not rely on any assumptions. To endow the building blocks and the representation itself with a causal interpretation requires some assumption on, for example, the assignment mechanism.
 $\square$
\end{comments}

\begin{comments}\rm
The lemma implies that the DID estimator  has an interpretation as a weighted average of simple estimators for the causal effect of changes in adoption dates, the $ \hat\tau_{t, aa'}$. Moreover, the estimator can be written as the sum  of three averages of these $ \hat\tau_{t, aa'}$. The first is a weighted average of the $\hat\tau_{t,\infty1}$, which are all averages of switching from never adopting to adopting in the first period, meaning that these are averages of changes in adoption dates that involve switching from not being treated at time $t$ to being treated at time $t$. The sum of the weights for these averages is one, although not all the weights are necessarily non-negative. The second sum is a weighted sum of 
 $ \hat\tau_{t, \infty a}$, for $a>t$, so that the causal effect always involves changing the adoption date from never adopting to adopting some time after $t$, meaning that the comparison is between potential outcomes neither of which involves being treated at the time. The sum of the  weights for these averages is one again.
 The third sum is a weighted sum of 
 $ \hat\tau_{t,  a1}$, for $a\leq t$, so that the causal effect always involves changing the adoption date from adopting prior to, or at time, $t$ relative to adopting at the initial time, meaning that the comparison is between potential outcomes both of which involves being treated at the time. These weights sum to minus one.
$\square$
\end{comments}

If we are willing to make the random adoption date assumption we can give this representation a causal interpretation:
\begin{theorem}\label{theorem1}
Suppose Assumption \ref{assumption_completely_random} holds.  
 Then $(i)$:
\[ \mme\left[\hat\tau_{t,aa'}\right]=\tau_{t,aa'},\]
and $(ii)$
\[ \mme\left[\htdid\right]
=\sum_{t\in\mmt} \gamma_{t,\plus} \tauinfone
+\sum_{t\in\mmt}\sum_{a> t} \gamma_{t,a} \tau_{t,\infty a}
-\sum_{t\in\mmt}\sum_{a\leq t} \gamma_{t,a} \tau_{t,a 1}.
\]
Suppose also Assumption \ref{assumption_no_anticipation} holds.
Then $(iii)$:
\[ \mme\left[\hat\tau_\did\right]=
\sum_{t\in\mmt}\sum_{a\leq t} \gamma_{t,a} \tau_{t,\infty a}.
\]
Suppose also Assumption \ref{assumption_exposure} holds. 
Then $(iv)$:
\[ \mme\left[\hat\tau_\did\right]=
\sum_{t=1}^T \gamma_{t,\plus}
\tauinfone.
\]
Suppose also Assumption \ref{assumption_constant_treatment_effect_time} holds. Then $(v)$:
\[ \mme\left[\hat\tau_\did\right]=
\tau_{\infone}.
\]
\end{theorem}

Part $(iii)$ of the theorem where we make the no-anticipation assumption is closely related to one of the results in 
 \citet{abraham2018estimating}, who make  a super-population common trend assumption that, in the super-population context, weakens our random adoption date assumption.
Part $(iv)$ of the theorem, where we assume both the exclusion restrictions so that the treatment is effectively a binary one, is related to the results in \citet{de2018two}, although unlike those authors we do not restrict the trends in the potential outcomes.

Without either Assumptions  \ref{assumption_no_anticipation}
or  \ref{assumption_exposure}, the estimand $\tau_\did$ has a causal interpretation, but it is not clear it is a very interesting one concerning the receipt of the treatment. With the no-anticipation assumption
(Assumption \ref{assumption_no_anticipation}), the interpretation, as given in part $(iii)$ of the theorem, is substantially more interesting. Now the estimand is a weighted average of $\tau_{t,\infty a}$ for $a\leq t$, with weights summing to one. These
$\tau_{t,\infty a}$ are the average causal effect of changing the adoption date from never adopting to some adoption date prior to, or equal to, time $t$, so that the average always involves switching from not being exposed to the treatment to being exposed to the treatment.

\subsection{The Randomization Variance of the  Difference-In-Differences Estimators}

In this section we derive the randomization variance for $\htdid$ under the randomized adoption date assumption. We do not rely on other assumptions here, although they may be required for making the estimand a substantively interesting one.
The starting point is the representation $\htdid=\sum_{t,a}\gamma_{t,a}\oy_{t,a}$. Because under Assumption 
\ref{assumption_completely_random} the weights $\gamma_{t,a}$ are fixed, the variance is
\[ \mmv(\htdid)=\sum_{t,a} \gamma^2_{t,a} \mmv(\oy_{t,a})+\sum_{(t,a)\neq (t',a')}\gamma_{t,a}\gamma_{t',a'} \mmc(\oy_{t,a},\oy_{t',a'}).\]
Note that the $\gamma_{t,a}$ are known.
Working out the variance $\mmv(\oy_{t,a})$, and finding an unbiased estimator for it, is straightforward. It is more challenging to infer the covariance terms $\mmc(\oy_{t,a},\oy_{t',a'}),$ and even more difficult to estimate them. In general that is not possible.
Note that for a sampling-based variance the $\gamma_{t,a}$ are not fixed, because in different samples the fractions with a particular adoption date will be stochastic. This in general leads to a larger variance, as we  verify in the simulations.

Define
\[ Y_{i}(a)=\sum_{t=1}^T \gamma_{t,a} Y_{it}(a),\hskip1cm
\oy(a)=\sum_{t=1}^T \gamma_{t,a} \oy_{t}(a)
\hskip1cm
{\rm and}\ \ 
\oy_a=\sum_{t=1}^T \gamma_{t,a} \oy_{t,a}
.\]
Now we can write $\hat\tau^\did$ as
\[ \hat\tau^\did=\sum_{a\in\mma}\sum_{t\in\mmt} \gamma_{t,a}\oy_{t,a}
=\sum_{a\in\mma} \oy_a.\]
Define also
\[ S^2_{a}=\frac{1}{N-1}\sum_{i=1}^N \left( Y_{i}(a)-\oy(a)\right)^2,\]
and
\[ V^2_{a,a'}=\frac{1}{N-1}\sum_{i=1}^N \left\{\left( Y_{i}(a)-\oy(a)\right)+\left( Y_{i}(a')-\oy(a')\right)\right\}^2.\]

\begin{theorem}\label{theorem2}
Suppose Assumptions \ref{assumption_completely_random} holds. Then the exact variance of $\htdid$ over the randomization distribution is
\[ \mmv\left(\htdid\right)=
\sum_{a\in\mma} S^2_{a}\left(\frac{1}{N_a}+\frac{T-1}{N}\right)
-\sum_{a\in\mma} \sum_{a'\in\mma,a'>a} \frac{V^2_{a,a'}}{N},
\]
with
\[  \mmv\left(\htdid\right)\leq \sum_{a\in\mma} S^2_{a}/N_a
.\]
\end{theorem}

\noindent{\bf Comment (ctd):} In our two period example with some units adopting in the second period and the others not at all, and $Y_{i1}(a)=0$, we have
\[ \gamma_1=\left(
\begin{array}{c}0 \\ 0
\end{array}\right),\hskip1cm 
 \gamma_2=\left(
\begin{array}{c}-1 \\1
\end{array}\right),\hskip1cm {\rm and}\ \ 
 \gamma_\infty=\left(
\begin{array}{c}1 \\ -1
\end{array}\right).\]
\[ S^2_{1}=0,\]
\[ S^2_{,2}=\frac{1}{N-1}\sum_{i=1}^N\left( Y_{i2}(2)-\frac{1}{N}\sum_{j=1}^N Y_{j2}(2)\right)^2,\]
\[ S^2_{\infty}=\frac{1}{N-1}\sum_{i=1}^N\left( Y_{i2}(\infty)-\frac{1}{N}\sum_{j=1}^N Y_{j2}(\infty)\right)^2,\]
\[ V^2_{1,2}=0, \hskip1cm S^2_{1,\infty}=0,\]
\[V^2_{2,\infty}=
 \frac{1}{N-1}
\sum_{i=1}^N \left(\left(Y_{i2}(2)-\frac{1}{N}\sum_{j=1}^N Y_{i2}(2)\right)
-\left(Y_{i2}(\infty)-\frac{1}{N}\sum_{j=1}^N Y_{i2}(\infty)\right)
\right)^2,\]
so that in this special 
\[ \mmv(\htdid)=
\frac{1}{N_2(N-1)}
\sum_{i=1}^N \left(Y_{i2}(2)-\frac{1}{N}\sum_{j=1}^N Y_{i2}(2)\right)^2\]
\[\hskip2cm + \frac{1}{N_\infty(N-1)}
\sum_{i=1}^N \left(Y_{i2}(\infty)-\frac{1}{N}\sum_{j=1}^N Y_{i2}(\infty)\right)^2\]
\[\hskip2cm - \frac{1}{N(N-1)}
\sum_{i=1}^N \left(\left(Y_{i2}(2)-\frac{1}{N}\sum_{j=1}^N Y_{i2}(2)\right)
-\left(Y_{i2}(\infty)-\frac{1}{N}\sum_{j=1}^N Y_{i2}(\infty)\right)
\right)^2
,\]
which agrees with the Neyman variance for a completely randomized experiment.
$\square$

\subsection{Estimating the Randomization Variance of the  Difference-In-Differences Estimators}

In this section we discuss estimating the variance of the DID estimator.
In general there is no unbiased estimator for $\mmv\left(\htdid\right)$. This is not surprising, because there is no such estimator for the simple difference in means estimator in a completely randomized experiment, and this corresponds to the special case with $T=1$. However, it turns out that just like in the simpled randomized experiment case, there is a conservative variance estimator. In the current case it is based on using unbiased estimators for the terms involving $S^2_{\gamma_a,a}$, and ignoring the terms involving
$V^2_{\gamma_a,a,\gamma_{a'},a'}$. Because the latter are non-negative, and enter with a minus sign, ignoring them leads to an upwardly biased variance estimator.
One difference with the simple randomized experiment case is that there is no simple case with constant treatment effects such that the variance estimator is unbiased.

Next, define the estimated variance of this by adoption date:
\[ s^2_{a}\equiv 
\frac{1}{N_a-1}\sum_{i:A_i=a} \left(Y_{i}
-\oy_{a}
\right)^2.\]
Now we can characterize the proposed variance estimator as
\[ \hvdid\equiv\sum_{a\in\mma} \frac{s^2_{a}}{N_a}.\]

\begin{theorem}\label{theorem3}
Suppose Assumption \ref{assumption_completely_random} holds. Then 
\[ \mme\left[\hvdid\right]\geq \mmv(\htdid),\]
so that $\hvdid$ is a conservative variance estimator for $\htdid$.
\end{theorem}

There are two important issues regarding this variance estimator. 
The first is its relation to the standard variance estimator for DID estimators. The second is whether  one can improve on this variance estimator given that in general it is conservative. 

The relevant variance estimators are the Liang-Zeger clustered variance estimator and the clustered bootstrap (\citet{Bertrand2004did, liang1986longitudinal}). Both have large sample justifications under random sampling from a large population, so they are in general not equal to the variance estimator here. 
In large samples both the Liang-Zeger and bootstrap variance will be more conservative than  $\hat\mmv_\did$ because they also take into account variation in the weights $\gamma_{t,a}$. These weights are kept fixed under the randomization scheme, because that keeps fixed the marginal distribution of the adoption dates. In contrast, under the Liang-Zeger calculations and the clustered bootstrap, the fraction of units with a particular adoption date varies, and that introduces additional uncertainty. 

The second issue is whether we can improve on the conservative variance estimator $\hat\mmv_\did$. In general there is only a limited ability to do so. Note, for example, that in the two period example this variance reduces to the  Neyman variance in randomized experiments. In that case we know we can improve on this variance a little bit exploiting heteroskedasticity, e.g.,  
 \citet{AGL14}, but in general those gains are modest.
 
\section{Some Simulations}

The goal is to compare the exact variance, and the corresponding estimator in the paper to the two leading alternatives, the Liang-Zeger (stata) clustered standard errors and the clustered bootstrap. We want to confirm settings where the proposed variance estimator differs from  the Liang-Zeger clustered variance, and settings where it is the same.
We have $N$ units, observed for $T$ time periods. We focus primarily on the case with $T=3$. The adoption date is randomly assigned, with $\pi_{I}=(\pi_1,\pi_2,\pi_3,\pi_\infty)=(0,0.67,0,0.33)$, and $\pi_{II}=(\pi_1,\pi_2,\pi_3,\pi_\infty)=(0,0.5,0.4,0.1)$.

We consider two designs for the potential outcome distributions in the population, the $Y_i(a)$ for $a\in\{1,2,3,\infty\}$.
In design A the potential outcomes,  are generated as
\[ \left(\begin{array}{c}Y_{i1}(2)\\
Y_{i1}(3)\\
Y_{i1}(\infty)\\
Y_{i2}(2)\\
Y_{i2}(3)\\
Y_{i2}(\infty)\\
Y_{i3}(2)\\
Y_{i3}(3)\\
Y_{i3}(\infty)\\
\end{array}\right)
\sim
{\cal N}
\left(
\left(\begin{array}{c}
0 \\0 \\0\\
4\\3\\3\\2\\2\\1
\end{array}
\right),\sigma^2
\left(\begin{array}{ccccccccc}
1 &  0  & 0 & 0  & 0 & 0 &0 & 0 & 0 \\
0 & 1   & 0 & 0  & 0 & 0 &0 & 0 & 0 \\
0 &  0  & 1 & 0  & 0 & 0 &0 & 0 & 0 \\
0 &  0  & 0 & 1  & 0 & 0 &0 & 0 & 0 \\
0 &  0  & 0 & 0  & 1 & 0 &0 & 0 & 0 \\
0 &  0  & 0 & 0  & 0 & 1 &0 & 0 & 0 \\
0 &  0  & 0 & 0  & 0 & 0 &1 & 0 & 0 \\
0 &  0  & 0 & 0  & 0 & 0 &0 & 1 & 0 \\
0 &  0  & 0 & 0  & 0 & 0 &0 & 0 & 1 
\end{array}
\right)
\right) .
\]

In this design the treatment effect is constant, and depends only on whether the adoption date preceeds the potential outcome date, or
\[ Y_{it}(a)={\bf 1}_{a\leq t}+\varepsilon_{it},\]
where the $\varepsilon_{it}$ are correlated over time.

In design B the potential outcomes  are generated as

\[ \left(\begin{array}{c}Y_{i1}(2)\\
Y_{i1}(3)\\
Y_{i1}(\infty)\\
Y_{i2}(2)\\
Y_{i2}(3)\\
Y_{i2}(\infty)\\
Y_{i3}(2)\\
Y_{i3}(3)\\
Y_{i3}(\infty)\\
\end{array}\right)
\sim
{\cal N}
\left(
\left(\begin{array}{c}
0 \\0 \\0\\
2\\1\\1\\2\\11\\1
\end{array}
\right),\sigma^2
\left(\begin{array}{ccccccccc}
1 &  0  & 0 & 0  & 0 & 0 &0 & 0 & 0 \\
0 & 10  & 0 & 0  & 0 & 0 &0 & 0 & 0 \\
0 &  0  & 1 & 0  & 0 & 0 &0 & 0 & 0 \\
0 &  0  & 0 & 1  & 0 & 0 &0 & 0 & 0 \\
0 &  0  & 0 & 0  & 1 & 0 &0 & 0 & 0 \\
0 &  0  & 0 & 0  & 0 & 1 &0 & 0 & 0 \\
0 &  0  & 0 & 0  & 0 & 0 &1 & 0 & 0 \\
0 &  0  & 0 & 0  & 0 & 0 &0 & 1 & 0 \\
0 &  0  & 0 & 0  & 0 & 0 &0 & 0 & 1 
\end{array}
\right)
\right) .
\]

Here the treatment effects depend on the treatment having been adopted, but the effect differs by the adoption date.

In design C the potential outcomes  are generated with positive correlations between the potential outcomes as

\[ \left(\begin{array}{c}Y_{i1}(2)\\
Y_{i1}(3)\\
Y_{i1}(\infty)\\
Y_{i2}(2)\\
Y_{i2}(3)\\
Y_{i2}(\infty)\\
Y_{i3}(2)\\
Y_{i3}(3)\\
Y_{i3}(\infty)\\
\end{array}\right)
\sim
{\cal N}
\left(
\left(\begin{array}{c}
0 \\0 \\0\\
2\\1\\1\\2\\11\\1
\end{array}
\right),\sigma^2
\left(\begin{array}{ccccccccc}
1 &  0.9  & 0.9 & 0  & 0 & 0 &0 & 0 & 0 \\
0.9 & 1  & 0.9 & 0  & 0 & 0 &0 & 0 & 0 \\
0.9 &  0.9  & 1 & 0  & 0 & 0 &0 & 0 & 0 \\
0 &  0  & 0 & 1  & 0.9 & 0.9 &0 & 0 & 0 \\
0 &  0  & 0 & 0.9  & 1 & 0.9 &0 & 0 & 0 \\
0 &  0  & 0 & 0.9  & 0.9 & 1 &0 & 0 & 0 \\
0 &  0  & 0 & 0  & 0 & 0 &1 & 0.9 & 0.9 \\
0 &  0  & 0 & 0  & 0 & 0 &0.9 & 1 & 0.9 \\
0 &  0  & 0 & 0  & 0 & 0 &0.9 & 0.9 & 1 
\end{array}
\right)
\right) .
\]

In design D the potential outcomes  are generated with negative correlations between the potential outcomes as

\[ \left(\begin{array}{c}Y_{i1}(2)\\
Y_{i1}(3)\\
Y_{i1}(\infty)\\
Y_{i2}(2)\\
Y_{i2}(3)\\
Y_{i2}(\infty)\\
Y_{i3}(2)\\
Y_{i3}(3)\\
Y_{i3}(\infty)\\
\end{array}\right)
\sim
{\cal N}
\left(
\left(\begin{array}{c}
0 \\0 \\0\\
2\\1\\1\\2\\11\\1
\end{array}
\right),\sigma^2
\left(\begin{array}{ccccccccc}
1 &  -0.4  & -0.4 & 0  & 0 & 0 &0 & 0 & 0 \\
-0.4 & 1  & -0.4 & 0  & 0 & 0 &0 & 0 & 0 \\
-0.4 & -0.4  & 1 & 0  & 0 & 0 &0 & 0 & 0 \\
0 &  0  & 0 & 1  & -0.4 & -0.4 &0 & 0 & 0 \\
0 &  0  & 0 & -0.4  & 1 & -0.4 &0 & 0 & 0 \\
0 &  0  & 0 & -0.4  & -0.4 & 1 &0 & 0 & 0 \\
0 &  0  & 0 & 0  & 0 & 0 &1 & -0.4 & -0.4 \\
0 &  0  & 0 & 0  & 0 & 0 &-0.4 & 1 & -0.4 \\
0 &  0  & 0 & 0  & 0 & 0 &-0.4 & -0.4 & 1 
\end{array}
\right)
\right) .
\]

For a particular design, eg $(A,2)$ draw the four  sets of three-component vectors of potential outcomes for each unit (the three components corresponding to the three time periods), one set for each of the values of $a\in\{1,2,3,\infty\}$. 
We keep these sets of potential outcomes  fixed across all simulations for a given design.
Then for each simulation draw the adoption date according to the distribution for that design, keeping the fraction of units with a particular adoption date fixed.

We want to look at variances and the corresponding confidence intervals based on four methods for estimating the variance for the DID estimator. The confidence intervals are Normal-distribution based, simply equal to the point estimates plus and minus 1.96 times the square root of the variances. 
We can write $\htdid$  as a regression estimator with $NT$ observations, and $N+T$ regressors. Let  with $j=1,\ldots, NT$. For observation $j$, $T_j\in\{1,\ldots,T\}$ denotes the time period the observation is from, and $N_j\in\{1,\ldots,N\}$ denotes the unit is corresponds to.
Now let $Y_j=Y_{N_j,T_j}$ and $W_j=W_{N_j,T_j}$, so that the regression function can be written as
\[ Y_j=\mu+\sum_{n=1}^{N-1} \alpha_n\mathbf{1}_{N_j=n}+
\sum_{t=1}^{T-1} \beta_t\mathbf{1}_{T_j=t}+\tau W_j+\varepsilon_j= Y_j=X_j^\top\theta+\varepsilon_j,\]
where
$X_j=(1,\mbf_{N_j=1},\ldots,\mbf_{N_j=N-1},,\mbf_{T_j=1},\ldots,\mbf_{T_j=T-1},W_j)$, and $\theta=(\mu,\alpha_1,\ldots,\alpha_{N-1},\beta_1,\ldots,\beta_{T-1},\tau)$.

We compare five variances. The first is exact randomization-based variance,
\[ \mmv_\did=\mmv\left(\hat\tau\right)=
\sum_{a\in\mma} \frac{S^2_{\gamma_a,a}}{N_a}
-\sum_{a\in\mma} \sum_{a'\in\mma,a'>a} \frac{V^2_{\gamma_a,a,\gamma_{a'},a'}}{N}.
\]
The other four  are estimators of the variance.

First, the feasible conservative variance estimator $\hvdid$.

Second, the standard Liang-Zeger clustered variance.
 Start with the representation $Y_j=X_j^\top\theta+\varepsilon_j$.
Let $\hat\varepsilon_j=Y_j-X_j^\top\hat\theta$ be the residual from this regression.
Calculate the variance as
\[ \widehat\mmv_{\rm LZ}=\left(\sum_{j=1}^J X_j X_j^\top\right)^{-1} 
\left(\sum_{n=1}^N \left(\sum_{j:N_j=n} X_j \hat\varepsilon_j\right)\left(\sum_{j:N_j=n} X_j \hat\varepsilon_j\right)^\top \right)
\left(\sum_{j=1}^J X_j X_j^\top\right)^{-1} ,
\]
and get the corresponding variance estimator for $\htdid$.

Third, the clustered bootstrap, $\widehat{\mmv}_{B1}$. Draw bootstrap samples based on drawing units, with all time periods for each unit drawn. Note that this explicitly changes from bootstrap sample to bootstrap sample the fraction of units with a particular adoption date.

Fourth,  a modification of the clustered bootstrap,  $\widehat{\mmv}_{B2}$, where we fix the fraction of units with each value for the adoption date.

In Table \ref{tabel_sim}  we report the results. For each  of the five variances we report the average of variance, and the coverage rate for the 95\% confidence interval.

\begin{table}[ht]
\centering
\caption{\textsc{: Simulations}}
\begin{tabular}{llrrrrrrrrrrr}
  \hline
Design & $\pi$ & N & $\mathbb{V}_{\text{did}}$ & Cov & $\hat{\mathbb{V}}_{\text{did}}$ & Cov & $\hat{\mathbb{V}}_{\text{LZ}}$ & Cov & 
$\hat{\mathbb{V}}_{\text{B1}}$ & Cov & $\hat{\mathbb{V}}_{\text{B2}}$ & Cov\\ 
  \hline
A & I &   30 & 0.144 & 0.951 & 0.239 & 0.979 & 0.214 & 0.974 & 0.232 & 0.975 & 0.219 & 0.973 \\ 
  B & I &   30 & 0.111 & 0.947 & 0.187 & 0.986 & 0.163 & 0.978 & 0.182 & 0.982 & 0.172 & 0.978 \\ 
  C & I &   30 & 0.201 & 0.953 & 0.217 & 0.947 & 0.181 & 0.925 & 0.211 & 0.942 & 0.200 & 0.932 \\ 
  D & I &   30 & 0.064 & 0.949 & 0.265 & 1.000 & 0.230 & 0.999 & 0.257 & 1.000 & 0.244 & 0.999 \\ 
  A & II &   30 & 0.112 & 0.946 & 0.165 & 0.972 & 0.146 & 0.966 & 0.158 & 0.969 & 0.142 & 0.956 \\ 
  B & II &   30 & 0.085 & 0.947 & 0.139 & 0.973 & 0.268 & 0.999 & 0.269 & 0.999 & 0.119 & 0.962 \\ 
  C & II &   30 & 0.184 & 0.949 & 0.191 & 0.939 & 0.279 & 0.983 & 0.285 & 0.981 & 0.162 & 0.920 \\ 
  D & II &   30 & 0.081 & 0.950 & 0.164 & 0.992 & 0.285 & 1.000 & 0.280 & 0.999 & 0.142 & 0.987 \\ 
  A & I &  150 & 0.027 & 0.953 & 0.047 & 0.991 & 0.045 & 0.989 & 0.047 & 0.989 & 0.046 & 0.989 \\ 
  B & I &  150 & 0.022 & 0.955 & 0.041 & 0.994 & 0.039 & 0.992 & 0.041 & 0.992 & 0.041 & 0.992 \\ 
  C & I &  150 & 0.035 & 0.956 & 0.038 & 0.960 & 0.036 & 0.956 & 0.037 & 0.955 & 0.037 & 0.954 \\ 
  D & I &  150 & 0.019 & 0.950 & 0.044 & 0.997 & 0.044 & 0.997 & 0.044 & 0.996 & 0.043 & 0.995 \\ 
  A & II &  150 & 0.020 & 0.952 & 0.033 & 0.989 & 0.033 & 0.989 & 0.033 & 0.987 & 0.032 & 0.987 \\ 
  B & II &  150 & 0.021 & 0.945 & 0.036 & 0.985 & 0.053 & 0.997 & 0.052 & 0.997 & 0.035 & 0.984 \\ 
  C & II &  150 & 0.034 & 0.952 & 0.035 & 0.953 & 0.051 & 0.985 & 0.052 & 0.983 & 0.034 & 0.947 \\ 
  D & II &  150 & 0.016 & 0.950 & 0.028 & 0.990 & 0.044 & 0.998 & 0.044 & 0.998 & 0.028 & 0.987 \\ 
   \hline 
\label{tabel_sim}
\end{tabular}
\end{table}

We see that the standard Liang-Zeger and the clustered bootstrap ($\widehat{\mmv}_{B1}$) substantially over-estimate the variance in Design B. The fixed adoption date bootstrap ($\widehat{\mmv}_{B2}$) and the proposed variance estimator ($\widehat{\mmv}_\did$) have the appropriate coverage.

\section{Conclusion}

We develop a design-based approach to Difference-In-Differences estimation in a setting with staggered adoption. We characterize what the standard DID estimator is estimating under a random adoption date assumption, and what the variance of the standard estimator is. We show that the standard DID estimatand is a weighted average of different types of causal effects, for example, the effect of changing from never adopting to adopting in the first period, or changing from never adopting to adopting later. In this approach the standard Liang-Zeger and clustered bootstrap variance estimators are unnecessarily conservative, and we propose an improved variance estimator.

\newpage 

\appendix

\centerline{\sc Appendix}

{\bf Proof of Lemma \ref{lemma:norestr}:} 
Let $\byp$ denote the $N\times (T\cdot (T+1))$ dimensional matrix with all the potential outcomes.
Because the units are exchangeable we can write the joint distribution of the potential outcomes and $\ba$ as
\[ f(\byp,\ba)=\prod_{i=1}^N f(\bypi,A_i).\]
Now we shall construct a distribution $f(\bypi,A_i)$ that satisfies two conditions. First, $A_i$ is independent of all the potential outcomes and second, the implied distribution for the adoption date and the realized outcome is consistent with the actual distribution. To do so we assume independence of the sets potential outcomes 
$Y_{i1}(a),\ldots,f_{iT}(a)$ for different $a$, and assume that
\[ f(Y_{i1}(a),\ldots,f_{iT}(a))=f(Y_{i1}(a),\ldots,f_{iT}(a)|A_i=a)=f(Y_{i1},\ldots,f_{iT}|A_i=a).\]
$\square$

{\bf Proof of Lemma \ref{lemma_binary_treatment}:} By Assumption  \ref{assumption_no_anticipation} we have
\[ Y_{it}(a)= \bfe _{a\leq t} Y_{it}(a)+ \bfe _{a>t} Y_{it}(\infty),\]
and by Assumption   \ref{assumption_exposure} we have
\[ Y_{it}(a)= \bfe _{a\leq t} Y_{it}(1)+ \bfe _{a>t} Y_{it}(a).\]
Combining the two assumptions implies
\[ Y_{it}(a)= \bfe _{a\leq t} Y_{it}(1)+ \bfe _{a>t} Y_{it}(\infty).\]
Hence
\[ Y_{it}(a')-Y_{it}(a)=
\bfe _{a'\leq t} Y_{it}(1)+ \bfe _{a'>t} Y_{it}(\infty)
-\left(\bfe _{a\leq t} Y_{it}(1)+ \bfe _{a>t} Y_{it}(\infty)\right)\]
\[\hskip1cm =\bfe _{a'\leq t<t} \left(Y_{it}(1)-Y_{it}(\infty)\right),\]
which proves part $(i)$.

For part $(ii)$
\[  \tau_{t,aa'}=\frac{1}{N}\sum_{i=1}^N \Bigl(Y_{it}(a')-Y_{it}(a)\Bigr)\]
\[\hskip1cm 
=\frac{1}{N}\sum_{i=1}^N \bfe _{a'\leq t<t} \left(Y_{it}(1)-Y_{it}(\infty)\right)
\]
\[\hskip1cm 
= \bfe _{a'\leq t<t} \frac{1}{N}\sum_{i=1}^N\left(Y_{it}(1)-Y_{it}(\infty)\right)
= \bfe _{a\leq t<a'}\tau_{t,\infty 1}
.\]
$\square$

{\bf Proof of Lemma \ref{lemma:testable}:} Part $(i)$ follows directly from the fact that the exclusion restrictions place restrictions only on potential outcomes that cannot be observed together.

Let us turn to part $(ii)$. By assumption
\[ Y_{it}(a)\ \indep\  A_i,\]
which as a special case includes
\[ Y_{i1}(\infty)\ \indep\  A_i.\]
Hence
\[ Y_{i1}(\infty)\ \indep\  A_i\ \Bigl|\ A_i\geq 2\]
which implies
\[ Y_{i1}\ \indep\  A_i\ \Bigl|\ A_i\geq 2\]
and thus
\[ Y_{i1}\ \indep\  A_i\ \Bigl|\ A_i\in\{2,\infty\},\]
which is a testable restriction.
$\square$

{\bf Proof of Lemma \ref{lemma_did}:} By Assumptions  \ref{assumption_no_anticipation} and \ref{assumption_exposure} we have
\[ Y_{it}(a)-Y_{it}(\infty)=\bfe _{a\leq t}\Bigl(Y_{it}(1)-Y_{it}(\infty)\Bigr).\]
By Assumptions \ref{assumption_constant_treatment_effect} and \ref{assumption_constant_treatment_effect_time}, $Y_{it}(1)-Y_{it}(\infty)=\tau_{1\infty}$, so that
\[ Y_{it}(a)-Y_{it}(\infty)=\bfe _{a\leq t}\tau_{1\infty}.\]
$\square$


{\bf Proof of Lemma \ref{lemma_did1}:}
Using the definition for $g(t,a)$, we can write $\htdid$ as
\[ \htdid=\frac{\sum_{i,t} \cwit  Y_{it} }{\sum_{i,t}\cwit^2}
=\frac{\sum_{t\in\mmt}\sum_{a\in\mma}\sum_{i:A_i=a}\cwit  Y_{it}  }{
N\sum_{t\in\mmt}\sum_{a\in\mma} \pi_a g(a,t)^2
}
=\frac{\sum_{t\in\mmt}\sum_{a\in\mma}\sum_{i:A_i=a}g(a,t)  Y_{it}  }{
N\sum_{t\in\mmt}\sum_{a\in\mma} \pi_a g(a,t)^2
}
\]
\[\hskip1cm
=\frac{\sum_{t\in\mmt}\sum_{a\in\mma}\sum_{i:A_i=a}g(a,t) N_a 
\oy_{t,a}  }
{
N \sum_{t\in\mmt}\sum_{a\in\mma} \pi_a  g(a,t)^2}
\]
\[\hskip1cm
=\frac{\sum_{t\in\mmt}\sum_{a\in\mma}g(a,t) \pi_a 
\oy_{t,a}  }
{
\sum_{t\in\mmt}\sum_{a\in\mma} \pi_a  g(a,t)^2}=\sum_{t,a} \gamma_{t,a} \oy_{t,a},
\]
where $\gamma_{t,a}$ is as given in (\ref{eq:did1}).
$\square$

{\bf Proof of Theorem  \ref{theorem1}:} First consider part $(i)$. We will show that
\[ \mme[\oy_{ta}]=\oy_t(a),\]
which in turn implies the result in $(i)$. We can write
\[ \mme[\oy_{ta}]=\mme\left[\frac{1}{N_a}\sum_{i=1}^N \bfe_{A_i=a} Y_{it}\right] =\mme\left[\frac{1}{N_a}\sum_{i=1}^N \bfe_{A_i=a} Y_{it}(a)\right].\]
By Assumption \ref{assumption_completely_random} this is equal to
\[\frac{1}{N_a}\sum_{i=1}^N \mme\left[\bfe_{A_i=a}\right] Y_{it}(a)=
\frac{1}{N_a}\sum_{i=1}^N \frac{N_a}{N} Y_{it}(a)=
\frac{1}{N}\sum_{i=1}^N  Y_{it}(a)=\oy_t(a)
,\]
which is the desired result.

Next consider part $(ii)$. By Lemma \ref{lemma_did1},
\[ \hat\tau_\did
=\sum_{t\in\mmt} \gamma_{t,\plus} \hat\tau_{t,\infty 1}
+\sum_{t\in\mmt}\sum_{a> t} \gamma_{t,a} \hat\tau_{t,\infty a}
-\sum_{t\in\mmt}\sum_{a\leq t} \gamma_{t,a} \hat\tau_{t,a 1},\]
so that
\[\mme\left[  \hat\tau_\did\right]
=\mme\left[\sum_{t\in\mmt} \gamma_{t,\plus} \hat\tau_{t,\infty 1}
+\sum_{t\in\mmt}\sum_{a> t} \gamma_{t,a} \hat\tau_{t,\infty a}
-\sum_{t\in\mmt}\sum_{a\leq t} \gamma_{t,a} \hat\tau_{t,a 1}\right],\]
which by Assumption \ref{assumption_completely_random} is equal to
\[
\sum_{t\in\mmt} \gamma_{t,\plus} \mme\left[\hat\tau_{t,\infty 1}\right]
+\sum_{t\in\mmt}\sum_{a> t} \gamma_{t,a}\mme\left[ \hat\tau_{t,\infty a}\right]
-\sum_{t\in\mmt}\sum_{a\leq t} \gamma_{t,a} \mme\left[\hat\tau_{t,a 1}\right].\]
This in turn, by part $(i)$, is equal to
\[
\sum_{t\in\mmt} \gamma_{t,\plus} \tau_{t,\infty 1}
+\sum_{t\in\mmt}\sum_{a> t} \gamma_{t,a}\tau_{t,\infty a}
-\sum_{t\in\mmt}\sum_{a\leq t} \gamma_{t,a} \tau_{t,a 1},\]
which finishes the proof of part $(ii)$.

Next consider part $(iii)$.
If Assumption \ref{assumption_no_anticipation}  holds, then for all $a> t$, $\tau_{t,\infty a}=0$, so that
\[\mme\left[  \hat\tau_\did\right]
=
\sum_{t\in\mmt} \gamma_{t,\plus} \tau_{t,\infty 1}
-\sum_{t\in\mmt}\sum_{a\leq t} \gamma_{t,a} \tau_{t,a 1}\]
\[\hskip1cm =\sum_{t\in\mmt}\sum_{a\leq t} \gamma_{t,a} \tau_{t,\infty a}.
\]

Next consider part $(iv)$. If also Assumption \ref{assumption_exposure} holds, then also  for all $a\leq t$, $\tau_{t,a 1}=0$, so that
\[\mme\left[  \hat\tau_\did\right]
=
\sum_{t\in\mmt} \gamma_{t,\plus} \tau_{t,\infty 1}
+\sum_{t\in\mmt}\sum_{a> t} \gamma_{t,a}\tau_{t,\infty a}
-\sum_{t\in\mmt}\sum_{a\leq t} \gamma_{t,a} \tau_{t,a 1}\]
\[\hskip1cm=
\sum_{t\in\mmt} \gamma_{t,\plus} \tau_{t,\infty 1}.\]

Finally, consider part $(v)$. This follows directly from part $(iv)$ in combination with the constant treatment effect assumption (Assumption \ref{assumption_constant_treatment_effect_time}).
$\square$

Next we give a preliminary result.
\begin{lemma}\label{multiple}
Suppose that Assumption  \ref{assumption_completely_random} holds. Then $(i)$ the variance of $\oy_a$ is 
\[ \mmv(\oy_a)=\frac{S^2_a}{N_a}\left(1-\frac{N_a}{N}\right),\]
(ii), the covariance of $\oy_a$ and $\oy_{a'}$ is
\[ \mmc(\oy_a,\oy_{a'})=-\frac{1}{2N}\left(S_a^2+S_{a'}^2-S^2_{aa'}\right)
=\frac{1}{2N}\left(S_a^2+S_{a'}^2-V^2_{aa'}\right),\]
 $(iii)$, the variance of the sum of the $\oy_a$ is
\[ \mmv\left(\sum_{a\in\mma} \oy_a\right)=
\sum_{a\in\mma}S_a^2\left(\frac{1}{N_a}+\frac{T-1}{N}\right)-\frac{1}{2N}\sumaap V^2_{aa'},\]
and
$(iv)$,
\[ \mmv\left(\sum_{a\in\mma}\oy_a\right)\leq 
\sum_{a\in\mma}\frac{S_a^2}{N_a}.\]
\end{lemma}
{\bf Proof of Lemma \ref{multiple}:}
Part $(i)$ follows directly from the variance of a sample average with random sampling from a finite population.

Next consider part $(ii)$. Define
\[ S^2_{aa'}=\frac{1}{N-1}\sum_{i=1}^N\left\{\left( Y_i(a')-\oy(a')\right)-
\left( Y_i(a)-\oy(a)\right)
\right\}.
\]
Recall that the variance of the difference between $\oy_{a'}$ and $\oy_a$ is
\[ \mmv(\oy_{a'}-\oy_a)=\frac{S^2_{a}}{N_a}+\frac{S^2_{a'}}{N_{a'}}-\frac{S^2_{aa'}}{N},\]
from the results in \citet{neyman1923, imbens2015causal} for completely randomized experiments with a binary treatment.
In general it is also true that 
\[ \mmv(\oy_{a'}-\oy_a)=\mmv(\oy_a)+\mmv(\oy_{a'})-2\mmc(\oy_a,\oy_{a'}).\]
Combining these two characterizations of the variance of the standard estimator for the average treatment effect, it follows that
the covariance is equal to
\[ \mmc(\oy_a,\oy_{a'})=\frac{1}{2}\left\{\mmv(\oy_a)+\mmv(\oy_{a'})- \mmv(\oy_{a'}-\oy_a)
\right\}\]
\[\hskip1cm =
\frac{1}{2}\left\{
\frac{S^2_a }{N_a}\left(1-\frac{N_a}{N}\right)
+
\frac{S^2_{a'} }{N_{a'}}\left(1-\frac{N_{a'}}{N}\right)
-\left\{\frac{S^2_{a}}{N_a}+\frac{S^2_{a'}}{N_{a'}}-\frac{S^2_{aa'}}{N}\right\}
\right\}\]
\[\hskip1cm =
-\frac{1}{2N}\left\{
S^2_a 
+
S^2_{a'}
-S^2_{aa'}
\right\}\]
\[\hskip1cm 
 =
-\frac{1}{2N}\left\{
S^2_a 
+
S^2_{a'}
+V^2_{aa'}-2S_a^2-2S_{a'}^2
\right\}
\]
\[\hskip1cm =
\frac{1}{2N}\left\{
S^2_a 
+
S^2_{a'}
-V^2_{aa'}
\right\}.
\]

Next, consider part $(iii)$. Using the result in part $(ii)$,
\[  \mmv\left(\sum_{a\in\mma} \oy_a\right)=
\sum_{a\in\mma} \mmv( \oy_a)+\sumaap \mmc(\oy_a,\oy_{a'})\]
\[\hskip1cm =\sum_{a\in\mma} \frac{S^2_a }{N_a}\left(1-\frac{N_a}{N}\right)
+\frac{1}{2N}\sumaap 
\left\{
S_a^2+S_{a'}^2-V^2_{aa'}
\right\}
\]
\[\hskip1cm =\sum_{a\in\mma} S^2_a \left(\frac{1}{N_a}-\frac{1}{N}+\frac{T}{N}\right)
-\frac{1}{2N}\sumaap 
V^2_{aa'}
\]
\[\hskip1cm =\sum_{a\in\mma}S_a^2\left(\frac{1}{N_a}+\frac{T-1}{N}\right)-\frac{1}{2N}\sumaap V^2_{aa'}.\]

Finally, consider part $(iv)$. The third term, the sum of $V^2_{aa'}$ terms is not directly estimable. Because it has a negative sign, we need to find a lower bound on this sum. A trivial lower bound is zero, but we can do better. We will show that 
\begin{equation}\label{peen} \frac{1}{2N}\sumaap 
V^2_{aa'}\geq \sum_{a\in\mma} S^2_a \frac{T-1}{N}.\end{equation}
This in turn implies
\[ -\frac{1}{2N}\sumaap
V^2_{aa'}\leq- \sum_{a\in\mma} S^2_a \frac{T-1}{N},\]
and thus
\[  \mmv\left(\sum_{a\in\mma} \oy_a\right)=\sum_{a\in\mma}S_a^2\left(\frac{1}{N_a}+\frac{T-1}{N}\right)-\frac{1}{2N}
\sum_{a\neq a'} V^2_{aa'}\]
\[\hskip1cm \leq 
\sum_{a\in\mma}S_a^2\left(\frac{1}{N_a}+\frac{T-1}{N}\right) -\sum_{a\in\mma} S^2_a \frac{T-1}{N}
\]
\[\hskip1cm =\sum_{a\in\mma}\frac{S_a^2}{N_a}.\]
The last inequality to prove is (\ref{peen}). First,
\[  V^2_{aa'}=
\frac{1}{N-1}\sum_{i=1}^N \left(\left(Y_i(a')-\oy(a')\right)+\left(Y_i(a)-\oy(a)\right)\right)^2\]
\[\hskip1cm =
\frac{1}{N-1}\sum_{i=1}^N \left\{\left(Y_i(a')-\oy(a')\right)^2+\left(Y_i(a)-\oy(a)\right)^2
+2\left(Y_i(a')-\oy(a')\right)\left(Y_i(a)-\oy(a)\right)^2
\right\}\]
\[\hskip1cm =\frac{1}{N}\left(S^2_a+S^2_{a'}+2\mmc(Y_i(a),Y_i(a'))\right).\]
Hence
\[  \frac{1}{2N}\sum_{a\neq a'} 
V^2_{aa'}=\frac{1}{2N}\sumaap 
\left\{S^2_a+S^2_{a'}+2\mmc(Y_i(a),Y_i(a'))\right\}\]
\begin{equation}\label{ptwee}\hskip1cm =\sum_{a\in\mma} S_a^2\frac{T}{N}+\frac{1}{N}\sum_{a\neq a'} \mmc(Y_i(a),Y_i(a')).\end{equation}
Next,
\[ 
0\leq \mmv\left(\sum_{a\in\mma} Y_i(a)\right)=\sum_{a\in\mma} \mmv(Y_i(a))+\sumaap \mmc(Y_i(a),Y_i(a')).\]
Therefore
\begin{equation}\label{pdrie} \sumaap \mmc(Y_i(a),Y_i(a'))\geq -\sum_{a\in\mma} \mmv(Y_i(a))=
-\sum_{a\in\mma} S^2_a
.\end{equation}
Combining (\ref{ptwee}) and (\ref{pdrie}) we get the bound
\[ \frac{1}{2N}\sumaap
V^2_{aa'} =\sum_{a\in\mma} S_a^2\frac{T}{N}+\frac{1}{N}\sumaap \mmc(Y_i(a),Y_i(a'))\]
\[\hskip1cm \geq \sum_{a\in\mma} S_a^2\frac{T}{N}-\sum_{a\in\mma} S^2_a=
\sum_{a\in\mma} S_a^2\frac{T-1}{N},
\]
which proves (\ref{peen}).
$\square$

{\bf Proof of Theorem  \ref{theorem2}:}  This follows directly from the results in  Lemma \ref{multiple}.
$\square$

{\bf Proof of Theorem  \ref{theorem3}:} By Assumption  \ref{assumption_completely_random} it follows that 
\[ \mme\left[ s^2_{\gamma_a,a}\right]=S^2_{\gamma_a,a}.\]
 This implies that
 \[ \mme\left[\hat\mmv_\did\right]
 =\mme\left[\sum_{a\in\mma} s^2_{\gamma_a,a}/N_a\right]
 =\sum_{a\in\mma} S^2_{\gamma_a,a}/N_a
 \geq  \mmv(\htdid),
 \] where the inequality is 
 by Theorem \ref{theorem2}.
 $\square$

\newpage

\bibliographystyle{plainnat}
\bibliography{references}

\end{document}